\documentstyle[emulateapj,psfig]{article}

\received{2 August 2001}  

\def\etal{{\it et al.~}}
\def\lsim{\hbox{ \rlap{\raise 0.425ex\hbox{$<$}}\lower 0.65ex\hbox{$\sim$} }}
\def\gsim{\hbox{ \rlap{\raise 0.425ex\hbox{$>$}}\lower 0.65ex\hbox{$\sim$} }}

\def\f(h{\hbox{$~\!\!^{\rm h}$}}

\def\ale{\mathrel{\hbox{\rlap{\hbox{\lower4pt\hbox{$\sim$}}}\hbox{$<$}}}}
\def\age{\mathrel{\hbox{\rlap{\hbox{\lower4pt\hbox{$\sim$}}}\hbox{$>$}}}}

\slugcomment{Submitted to The Astrophysical Journal Letters}

\lefthead{Djorgovski et al.} 
\righthead{Reionization Epoch}

\begin{document}

\title{On the Threshold of the Reionization Epoch
\footnotemark}

\footnotetext{Based on the observations obtained at the W.~M.~Keck Observatory
which is operated by the California Association for Research in Astronomy, a
scientific partnership among California Institute of Technology, the University
of California and the National Aeronautics and Space Administration.} 

\author{S. G. Djorgovski, S. Castro}

\affil{Palomar Observatory 105--24, California Institute of Technology,
       Pasadena, CA 91125, USA; {\tt [george,smc]@astro.caltech.edu}}

\author{D. Stern}

\affil{Jet Propulsion Laboratory, 
       Pasadena, CA 91109, USA; {\tt stern@zwolfkinder.jpl.nasa.gov}}

\author{A. A. Mahabal}

\affil{Palomar Observatory 105--24, California Institute of Technology,
       Pasadena, CA 91125, USA; {\tt aam@astro.caltech.edu}}


\begin{abstract}
Discovery of the cosmic reionization epoch would represent a significant
milestone in cosmology.  
We present Keck spectroscopy of the quasar SDSS 1044--0125, at $z = 5.73$.
The spectrum shows a dramatic increase in the optical depth at observed
wavelengths $\lambda \gsim 7550$ \AA, corresponding to $z_{abs} \gsim 5.2$. 
Only a few small, narrow transmission regions are present in the spectrum
beyond that point, and out to the redshifts where the quasar signal begins.
We interpret this result as a signature of the trailing edge of the cosmic
reionization epoch, which we estimate to occur around 
$\langle z \rangle \sim 6$ (as indeed confirmed by subsequent observations
by Becker \etal), and extending down to $z \sim 5.2$.
This behavior is expected in the modern theoretical models of the reionization
era, which predict a patchy and gradual onset of reionization. 
The remaining transmission windows we see may correspond to the individual
reionization bubbles (Str\"omgren spheres) embedded in a still largely neutral
intergalactic medium, intersected by the line of sight to the quasar. 
Future spectroscopic observations of quasars at comparable or larger redshifts
will provide a more detailed insight into the structure and extent of the
reionization era. 
\end{abstract}


\section{Introduction}

There has been a great progress over the past several years in our 
understanding of galaxy evolution and formation.  Samples of normal galaxies
are now studied out to $z \sim 4.5$ (\cite{stei99}), and several galaxies
are now known at $z > 5$ (see \cite{ss99} for a review and references).
Quasars at $z > 5$ (\cite{s+00}, \cite{zheng00}, \cite{fan00}, and references
therein) also represent a valuable probe of both galaxy and structure
formation, and the intervening primordial intergalactic medium (IGM). 

The observational frontier is now shifting to the formation of the first
objects, protogalaxies and primordial AGN, which is generally expected to occur
some time in the redshift interval $z \sim 6 - 15$ or so.  As the first sources
of UV radiation turn on, they reionize the universe, ending the ``dark ages''
which start at the recombination epoch ($z \sim 1100$). In this ``cosmic
renaissance'' (effectively, the start of the galaxy formation epoch) the
universe undergoes a phase transition from being neutral to being mostly
ionized.  

Detection of the reionization epoch would thus be a major cosmological
milestone.  The standard observational test is the prediction of an extended,
optically thick absorption due to neutral hydrogen at $\lambda_{rest} < 1216$
\AA\ (\cite{gp65}).  A limit to this effect at $z \approx 5$ was published by
\cite{shcm99}.  To date, only a gradual thickening of the absorption due to
the Ly$\alpha$ forest was seen.

In this Letter we present evidence which suggests that we are already probing 
the trailing end of the reionization era, at $z \sim 5.5 \pm 0.3$ or so.  
The evidence is based on the high S/N, Keck spectroscopy of the quasar 
SDSS 1044--0125 discovered by \cite{fan00}. 
Subsequent recent observations of quasars at $z \gsim 6$ by \cite{bec01}
provide an additional evidence that the reionization era indeed occurs
around $z \sim 6$.  Taken together, the available data support a picture of
an extended and patchy reionization era, ending at $z \sim 5 - 6$.

\section{Observations and Data Reductions}

Our low-resolution spectra were obtained on the W.~M.~Keck Observatory 10-m
telescope (Keck I) on UT 2000 December 30, using the Low Resolution Imaging
Spectrometer (LRIS; \cite{occ+95}).  The observations were obtained with the
400 lines mm$^{-1}$ grating ($\lambda_{\rm blaze} = 8500$\AA) through 1.2
arcsec slitlets, with two different slitmasks, with a mean dispersion of
$\approx 1.86$ \AA\ pixel$^{-1}$, and a GG495 long-pass order-sorting filter. 
The first set, totalling 2400~s of integration, was at a slit position angle PA
= 249.6$^\circ$, covering the quasar spectrum $\sim 6240$ \AA\ $-$ 1 $\mu$m,
and a mean airmass $\approx 1.08$.  The second, with 3600~s of integration, was
at PA = 137$^\circ$ (very close to parallactic angle at the time), covering
$\sim 7300$ \AA\ $-$ 1 $\mu$m, and a mean airmass $\approx 1.13$. 
The differential slit losses are estimated to be negligible for our purpose.
Individual integrations were dithered along the slit.  Data were reduced in 
IRAF, using standard slit spectroscopy procedures.  Ne+Ar arc lamp spectra
obtained through the masks were used for wavelength calibrations, and the
wavelength zeropoints adjusted using telluric emission lines.  The night was
photometric, but unfortunately no flux standards were observed with these
slitmask+grating combinations, and we used an average of archival response
curves for this grating obtained earlier.  Our spectroscopic magnitudes are in
an excellent agreement with the CCD photometry presented by \cite{fan00}. 

The combined LRIS spectrum is shown in Fig.~1.  It shows a dramatic drop
due to the Ly$\alpha$ absorption at $\lambda_{obs} \lsim 8100$ \AA, and a
second discontinuity at $\lambda_{obs} \lsim 6900$ \AA, due to the
Ly$\beta$ forest (this may be the strongest detection of the Ly$\beta$ 
drop observed to date).
In order to estimate the possible continuum level in the absorbed region,
we use three power-laws, $f_\nu \sim \nu^\alpha$, with $\alpha = [0,-0.5,-1]$,
which span a plausible range for quasars.  They have been normalized to
$f_\nu = 72~\mu$Jy at 9000 \AA, and are shown as dashed lines in Fig. 1.

Our high-resolution spectra were obtained on the W.~M.~Keck Observatory 10-m
telescope (Keck II) on UT 2000 April 28, UT 2001 January 1 and 2, and UT 2001
March 24, using the Echelle Spectrograph and Imager (ESI; \cite{mil00}). 
We used the Echelle mode which contains ten orders, with a complete optical
wavelength coverage, from $\sim 3900$ \AA\ to $\sim 10900$ \AA.  The instrument
has a spectral resolution of 11.4 km s$^{-1}$ pixel$^{-1}$, and a mean
dispersion in the wavelength region of interest here of $\approx 0.154$ \AA\
pixel$^{-1}$. A total of 11 exposures of 1800 s each were obtained, some in
slightly non-photometric conditions.  Data were reduced using standard
procedures. We used the program MAKEE (written by T. Barlow) to reduce the
spectra. Individual exposures from each night were combined prior to spectrum
extraction using a rejection algorithm to remove cosmic rays. Spectra were then
optimally extracted.  Exposures of bright stars were used to provide the
spectrum traces (necessary due to a heavy absorption present in the quasar
spectrum).  Dispersion solutions were found from exposures of arc lamps,
spectra were corrected to the Heliocentric system, and the wavelengths were
transformed to vacuum values.  The spectra for each night were flux calibrated
using a single response curve measured during one of the nights, and averaged
using the exposure time weighting. 

Since the flux zero-points for the ESI data are uncertain, we convolved both
ESI and LRIS spectra with Gaussians with $\sigma = 20$ \AA, thus bringing
them to effectively the same, very low resolution.  From the ratio of these
spectra we determined the flux correction factor, and applied it to the ESI
data.

The final ESI spectrum is shown in Fig.~2.  The absence of flux (save for a
few remaining narrow gaps in absorption) in the wavelength interval
$\sim 7550 - 8100$ \AA, between the patch of the Ly$\alpha$ forest in
the wavelength range $\sim 6900 - 7550$ \AA\ and the quasar signal at
$> 8100$ \AA, is quite striking.  (The Mg II doublet at $\lambda \approx 9180$
\AA\ was noted by \cite{fan00}; using a weighted average of several lines,
we measure the absorber redshift to be $z = 2.27865$.)

The sharp Ly$\beta$ drop at $\lambda \approx 6910$ \AA\ allows us to
estimate a better redshift for the quasar: $z = 5.73 \pm 0.01$.  This is less
than 5.80 originally estimated by \cite{fan00}, for the following reason. 
The object was found to be a BAL quasar by \cite{maio01}.  We believe that
its Ly$\alpha$ line is nearly completely absorbed, and that \cite{fan00}
mistakenly interpreted the red half of the [N V] 1240 line as Ly$\alpha$.
This unfortunately precludes the study of the red wing of the Ly$\alpha$
absorption, which may contain useful information about the structure of
the reionization front (\cite{mr00}, \cite{lb01}, \cite{bl01}).
Our redshift is supported by the possible Ly$\gamma$ line and a drop near
$\lambda \approx 6540$ \AA, and a Lyman limit (clearly seen in the 2-d spectra
images) at $\lambda \approx 6135$ \AA.
The effects of the blue BAL wing of [N V] and Ly$\alpha$ may extend as far
as $\lambda \sim 7900 - 8000$ \AA, but this is not critical for our discussion
below.

\section{Discussion and Conclusions}

There has been much recent progress in theoretical understanding and modeling
of the reionization era (excellent reviews include, e.g., \cite{mad00},
\cite{lb01}, \cite{bl01}, \cite{shap01}, etc.).  A good understanding of the
structure and extent of the reionization is important by itself, as it reflects
the earliest phases of structure formation, and also for the modeling of CMBR
foregrounds at high angular frequencies. 

A simple picture of a clean-cut Gunn-Peterson trough now appears unlikely.  
The key issue is the clumpiness of the IGM, and the gradual development and
clumpy distribution of the first ionizing sources, either protogalaxies or
early AGN (see, e.g., \cite{mehr00}).  The reionization is expected to occur 
gradually as the UV emissivity increases (cf. \cite{mcdme01}), and ionization
overcomes the recombination rate, with the lowest density regions becoming
fully reionized first.  This is also suggested by modern numerical simulations
(e.g., \cite{gned00}, \cite{cfgj00}, \cite{uns01}, etc.) which predict an
extended period of reionization, ranging from $z \sim 15$ to $z \sim 5$ or so. 

As we approach the reionization era from the lower redshifts, the Ly$\alpha$
forest thickens, with an occasional transmission gap due to the intersection of
ionized bubbles along the line of sight; eventually a complete Gunn-Peterson
trough is reached.  In other words, the inherent non-uniformity of galaxy and
structure formation is reflected in the structure of the IGM phase transition
corresponding to the reionization.  Further complications arise from the
proximity effect due to the source used to probe the IGM, i.e., a luminous
quasar, and the nature, luminosity, and duration of other sources near the
line of sight.

This general picture is illustrated well in Fig. 20 of \cite{lb01}
(which is the same as Fig. 40 of \cite{bl01}, or Fig. 6 of \cite{loeb99}).
The qualitative correspondence with the observed spectrum of SDSS 1044--0125
(Fig. 2) is striking.  The dramatic increase in the opacity of the 
Ly$\alpha$ forest at $\lambda \gsim 7550$ \AA, i.e., $z \gsim 5.2$, is
exactly what is expected in the approach to (or the tail end of) the 
reionization era.  This is further illustrated in Fig. 3, which shows a
dramatic thickening of the Ly$\alpha$ forest absorption at these redshifts.
A slightly different interpretation is that we are seeing a somewhat later
evolutionary stage of the reionization process, i.e., remaining islands of
(mostly) neutral gas embedded in a growing sea of ionized hydrogen.

The strong observed Ly$\beta$ break may also be due in part to the patches
of diffuse absorption we see so clearly in Ly$\alpha$.  The overall appearance
of the spectrum is suggestive of some of the models by \cite{hl99}, for the
reionization redshift (in their terminology) a few percent lower than the
source redshift.

The dark portions of our spectrum at $z \sim 5.2 - 5.6$ have the flux
consistent with zero, to within the photon noise.  The lower limit to the 
optical depth (r.m.s., per pixel) is $\tau \gsim 4$; if we average the flux
over the dark portions of the spectrum, this limit is considerably higher,
$\tau \gsim 6$ or 7, depending on the redshift window used.  Even if we assume
a very conservative systematic sky subtraction error of $\sim 1$\% of the
continuum, the implied optical depth limit would be $\tau \gsim 4.6$.
The extrapolation of the empirical scaling laws found by \cite{prs93}
and \cite{kcdo01} to these redshifts suggests $\tau \sim 2$.  This again
indicates that we are seeing more absorption than would be expected from a
simple extrapolation of the Ly$\alpha$ forest.

The few remaining transmission spikes are naturally interpreted as being
due to the as-yet unpercolated reionization bubbles along the line of sight.
An issue arises of whether the damping wings of the remaining neutral hydrogen
clouds would suppress such transmission spikes (cf. \cite{miresc98}).  However,
as shown by \cite{mr00} and \cite{ch00}, this depends strongly on the extent of
the Str\"omgren spheres produced by the ionized sources, i.e., their
luminosities and lifetimes, akin to the usual quasar proximity effect. 
Indeed, one expects some clustering of the first luminous sources, which
are expected to form at the highest peaks of the density field, due to
biasing (cf. \cite{djorg99} and references therein).  The probably massive
host of SDSS 1044--0125 is likely to have some luminous neighbors.

We also checked whether any of the dark regions we see at $z \sim 5.2 - 5.6$
may be DLA systems with associated metallic lines, mainly the C IV doublet.  
None were found, as is expected from as yet unenriched gas.

In their discovery paper, \cite{fan00} addressed the issue of reionization, and
concluded that it is not detected in their data, on the basis of the few
remaining transmission spikes.  Their were heavily binned spectrum seems roughly
comparable to our LRIS data shown in Fig. 1, from which we cannot conclude much;
a higher resolution spectrum, such as our spectrum shown in Fig. 2, is
necessary.  Furthermore, in their original redshift interpretation (which was
due to the then unknown BAL nature of the object) \cite{fan00} may have
mistaken the leftover quasar flux around the Ly$\alpha$ as being a part of the
Ly$\alpha$ forest. 

After this paper was submitted, \cite{bec01} presented spectroscopy of this and
three additional quasars at $z \sim 5.82 - 6.28$, discovered by \cite{fan01}. 
They present a compelling evidence for a Gunn-Peterson trough in the spectrum
of the most distant quasar, suggesting the reionization epoch at $z \sim 6$,
as anticipated here.  This is a crucial result.

However, the discussion of the remaining spectra by \cite{bec01} was limited by
the available data, with relatively short exposure times.  In order to increase
the apparent S/N ratio, they binned the spectra by a factor of $\sim 25$ in
wavelength, from $\sim 0.154$ \AA\ pixel$^{-1}$ to 4 \AA\ pixel$^{-1}$.  The
resulting loss of resolution makes it hard to detect dark windows similar to
those seen in our ESI spectrum, especially given the differences in the S/N
ratio.  They further evaluated the mean optical depth in very wide bins, with
$\Delta z = 0.2$, which clearly precludes the detection of any dark windows
with a smaller extent in redshift.  Given these differences in the data and the
analysis, we see no inconsistencies with our results. 

Taken together, the data so far suggest an extended and patchy end to the
reionization era, as expected from modern models of structure formation and
of the reionization itself.  Probing along more lines of sight with a high-S/N,
high-resolution spectroscopy is necessary in order to place more quantitative 
observational constraints.  Further insights will be obtained from direct
detections of luminous sources responsible for the reionization at these and
higher redshifts, their luminosity function and clustering properties. 

\acknowledgments

We want to thank the staff of the W.~M.~Keck Observatory for their expert
assistance.  The LRIS data were obtained in the course of a collaborative
project with F. Harrison and P. Mao.  We thank numerous colleagues whose
constructive comments helped us improve the discussion presented in the paper.
SGD acknowledges partial funding from the Bressler Foundation.  The work of DS
was carried out at the Jet Propulsion Laboratory, Caltech, under a contract
with NASA.



\begin{figure*}[tbp]
\centerline{\psfig{file=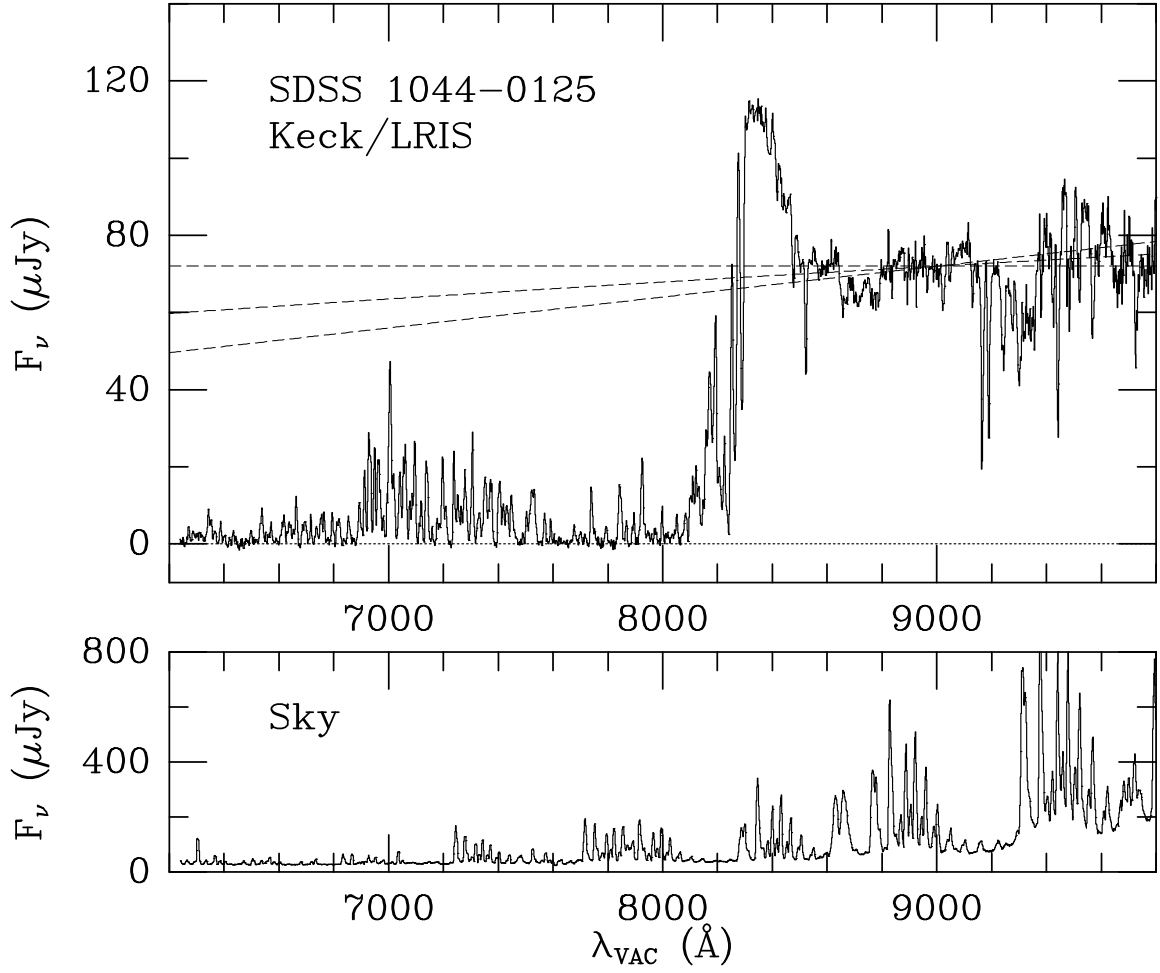,width=6.0in,angle=0}}
\caption[]{
Spectrum of SDSS 1044--0125 obtained with LRIS (top), and the corresponding
night sky (bottom).  The three dashed lines represent a plausible range of
the unabsorbed quasar power-law continua.
}
\label{fig:fig1}
\end{figure*}


\begin{figure*}[tbp]
\centerline{\psfig{file=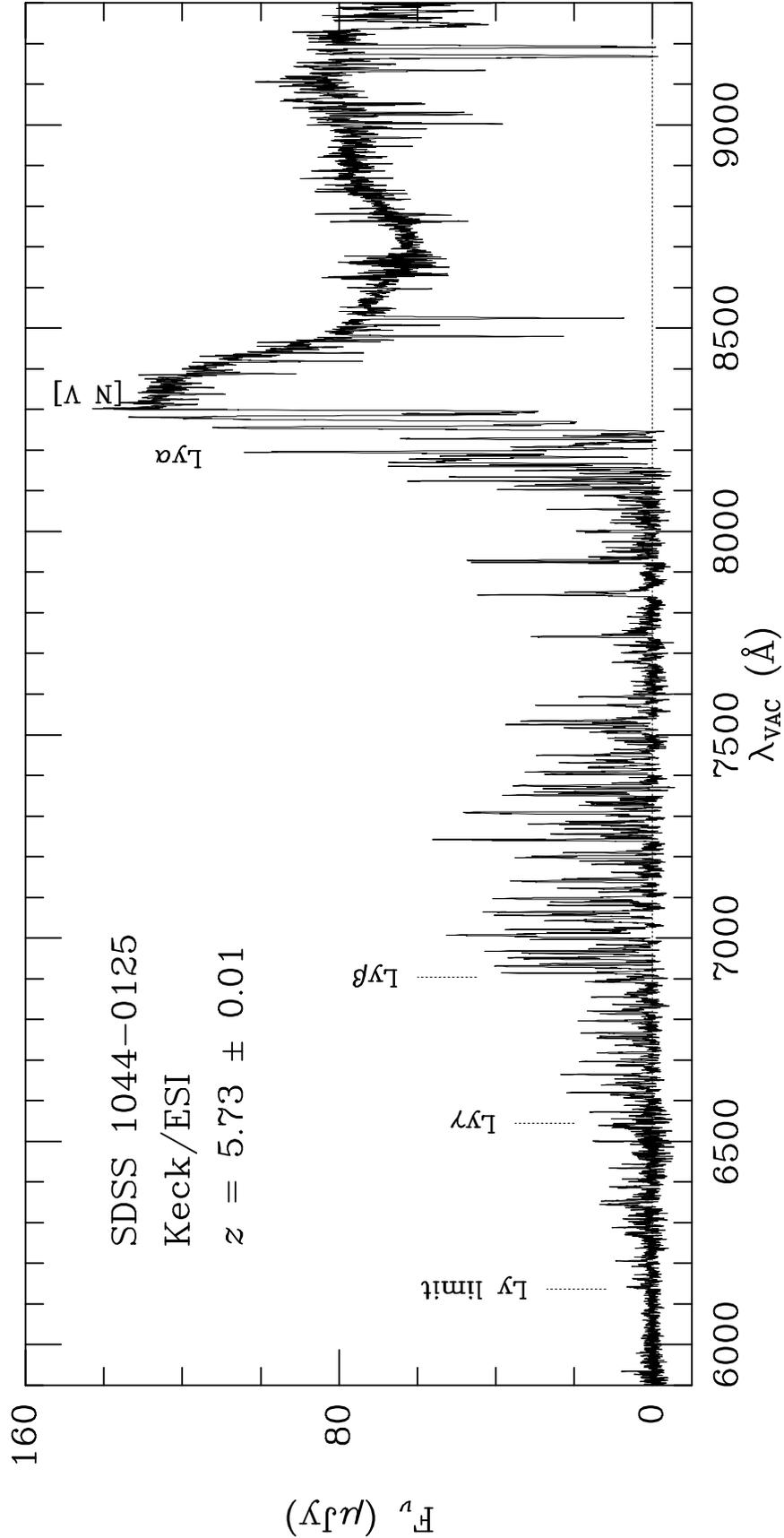,width=4.5in,angle=0}}
\caption[]{
Spectrum of SDSS 1044--0125 obtained with ESI.  Lyman series breaks 
corresponding to redshift $z = 5.73$ are indicated.  In this redshift
interpretation, most of the quasar's Ly$\alpha$ line is absorbed.
The dramatic change in the density of the Ly$\alpha$ forest at 
$\lambda \gsim 7550$ \AA, and probably the corresponding Ly$\beta$ 
absorption at $\lambda \lsim 6900$ \AA, are suggestive of the onset 
of the reionization era at $z > 5.2$.
}
\label{fig:fig2}
\end{figure*}


\begin{figure*}[tbp]
\centerline{\psfig{file=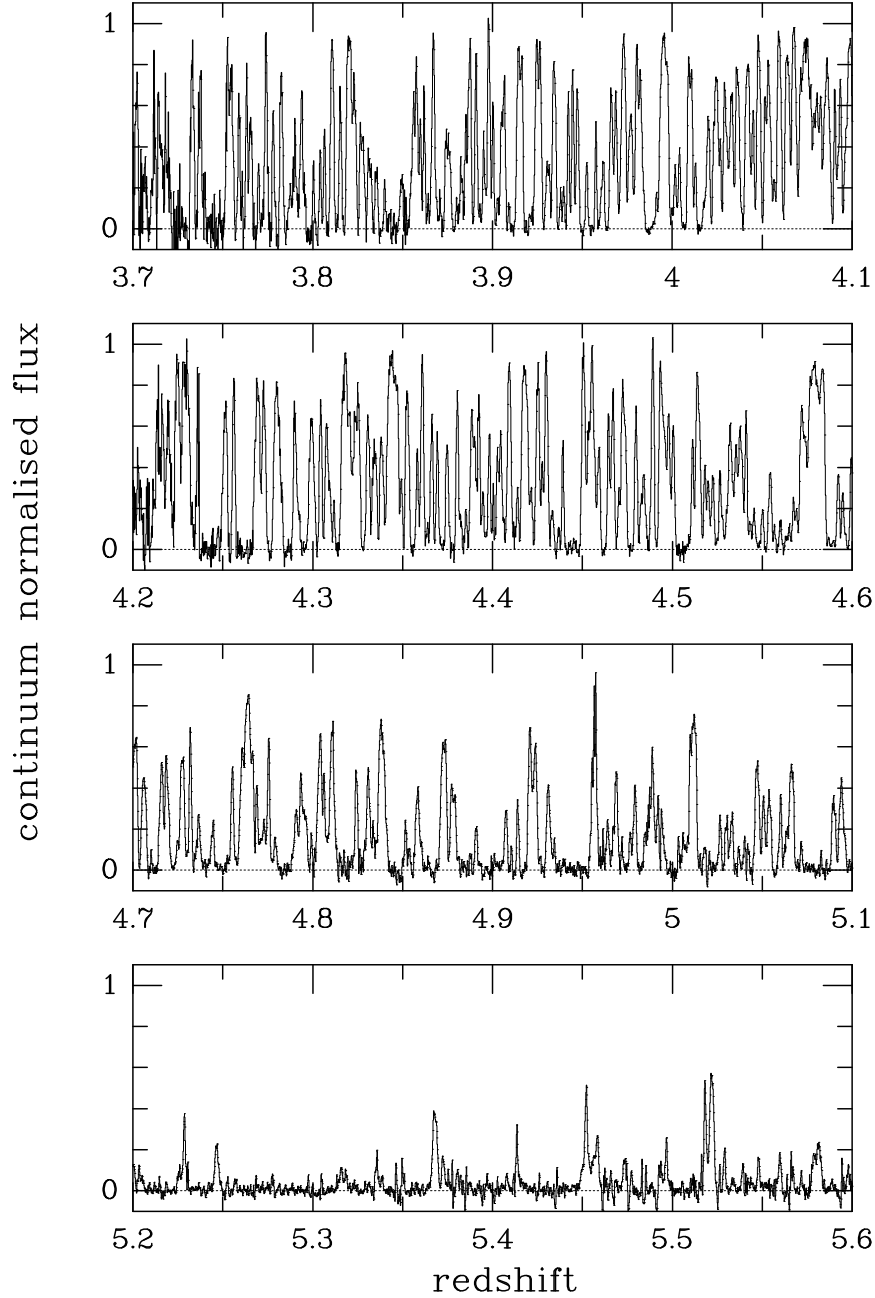,width=4.5in,angle=0}}
\caption[]{
A schematic illustration of the evolution of the Ly$\alpha$ forest, in four
redshift intervals.  The top two windows are from an ESI spectrum of 
SDSS 1737+5828 at $z = 4.94$; the bottom two are from the spectrum of SDSS
1044--0125, presented here.  The spectra have been renormalised by the best
estimate of the continuum (for the bottom two panels, we used the middle
power-law shown in Fig. 1).  Almost all narrow spikes in the bottom panel are
due to imperfect night sky emission lines subtraction. 
}
\label{fig:fig3}
\end{figure*}

\end{document}